\documentclass{iopart}
\usepackage{graphicx}

\usepackage{iopams}

\newcommand{\TSFT}{T_{\rm SFT}}
\newcommand{\Ap}{A_+}
\newcommand{\Ax}{A_{\times}}

\begin{document}


\title[Generalized PowerFlux methods ... periodic gravitational waves]{Using generalized PowerFlux methods to estimate the parameters of periodic gravitational waves}

\author{Gregory Mendell$^1$ and Karl Wette$^2$}

\address{$^1$LIGO Hanford Observatory, Richland, WA, USA}
\address{$^2$The Australian National University, Canberra, ACT, Australia}
\ead{gmendell@ligo-wa.caltech.edu}

\begin{abstract}
We investigate methods to estimate the parameters of the gravitational-wave signal from a spinning neutron star using Fourier transformed segments of the strain response from an interferometric detector. Estimating the parameters from the power, we find generalizations of the PowerFlux method. Using simulated elliptically polarized signals injected into Gaussian noise, we apply the generalized methods to estimate the squared amplitudes of the plus and cross polarizations (and, in the most general case, the polarization angle), and test the relative detection efficiencies of the various methods.
\end{abstract}



\section{Introduction: parameter estimation using power}

A periodic gravitational wave incident on an interferometric detector will produce a strain response of the form
\begin{equation}
\label{lhot}
h(t) = \Ap F_+(\psi,t) \cos\Phi(t) + \Ax F_\times(\psi,t) \sin\Phi(t),
\end{equation}
where $h(t)$ is the strain, $\Ap$ and $\Ax$ are the amplitudes of the plus and cross polarizations of the gravitational wave, $F_+$ and $F_\times$ are the respective sky-position dependent response functions (or antenna patterns) of the detector, $\psi$ is the polarization angle, and $\Phi$ is the gravitational wave phase,
which contains modulations from Doppler shifts due to the relative motion between
the source and the detector and the frequency evolution of the source \cite{jks}. (The response of a bar
detector can be written in a similar form, but with antenna patterns differing from those in \cite{jks}.)

The LIGO Scientific Collaboration (LSC) has used fully-coherent and semi-coherent methods to search
for unknown sources of these waves, while also developing hierarchical schemes that use a combination of these methods (see \cite{s2fstat, s4semicoherentpaper} and references therein). The starting procedure for these methods is to divide the strain data into short segments and take the discrete Fourier transform of each, to create Short Fourier Transforms (SFTs) of the data.
The time-baseline used to generate the SFTs is designated $\TSFT$. This is typically chosen
to be 30~min, so that signals with frequencies less than approximately 1000~Hz from an isolated source will not, within this time period, shift in frequency by more than half the width of an SFT frequency bin.  (The method can also be applied to sources in binary systems, by using a shorter time-baseline dependent on the size of the expected orbital Doppler shifts.)

The PowerFlux method \cite{pf} is a semi-coherent method for detecting periodic gravitational waves
using a weighted average of the power from SFTs. Compared with the StackSlide and weighted Hough methods, that also combine power from SFTs, in general the PowerFlux method has the highest sensitivity \cite{s4semicoherentpaper}. Hierarchical and fully-coherent searches that use a coherent time-baseline longer that 30~min are more sensitive, but also more computationally costly. Thus, PowerFlux remains a vital search method for periodic gravitational waves.

PowerFlux is able to estimate the squared amplitude (either $\Ap^2$ or $\Ax^2$) of a linearly polarized signal ($\Ax = 0$ or $\Ap = 0$) or a circularly polarized signal ($\Ap^2 = \Ax^2$), which can also be used as a detection statistic.
However, it does not directly estimate the parameters of elliptically polarized signals ($\Ap^2 \ne \Ax^2 \ne 0$ in general), nor does it directly recover the polarization angle $\psi$. In this paper we investigate whether it is possible to extract this extra information from a potential signal present in the SFTs and, if so, whether we can find a method using power from SFTs that is more efficient at detecting signals than standard PowerFlux.

We have previously (in \cite{mw}) investigated using the real and imaginary parts of the SFTs to estimate $\Ap$, $\Ax$, and $\psi$ using the method in \cite{jks}; the method worked well for searches with long coherent time-baselines, but failed to be robust on time-baselines of 30~min, for a number of reasons, as detailed in \cite{mw}. Here, we show instead how to estimate the parameters of elliptically polarized signals using the power in SFTs, thus generalizing the PowerFlux method.
As a first step, we present a new derivation of the PowerFlux method in the next section.  The subsequent sections give two generalizations, and then a comparision of the detections efficiencies of the methods (and the distributions of the estimated parameters). The last section gives another way to form a detection statistic using SFT power, analogous to finding a maximum likelihood statistic, and suggests future work.



\section{Derivation of the PowerFlux method}

By definition of $\TSFT$, we can treat $F_+$, $F_\times$, and the frequency of the signal at the detector
as constant over the duration of an SFT. Ignoring losses due to the difference between this frequency and the closest SFT bin frequency, the normalized power in the signal is
\begin{equation}
\label{pOneSFT}
{2 | \tilde{h} |^2 \over \TSFT } =
0.5 (\Ap^2F_+^2 + \Ax^2F_\times^2)\TSFT,
\end{equation}
where $\tilde{h}$ is the discrete Fourier transform of $h(t)$ divided by the sample rate of the data, and it is understood that $F_+$ and $F_\times$ are evaluated at the midpoint of each SFT. Equation~\eref{pOneSFT} represents the expected signal power for an elliptically polarized signal from one SFT. We label the SFTs using index $\alpha$, and consider searching for a linearly polarized signal (with $\Ax = 0$) which is present in the sequence of SFTs. Given the SFT data, $\tilde{x}_\alpha$, we define
\begin{equation}
\label{Palpha}
P_\alpha = { 2 | \tilde{x}_\alpha |^2 \over \TSFT }
\end{equation}
to be the power taken from the SFT bin closest to the expected signal frequency, and
\begin{equation}
\label{NWSSDP}
g = \sum_{\alpha} { [ P_\alpha - 0.5 \Ap^2 F_{+ \alpha}^2 \TSFT ]^2 \over S_\alpha^2 },
\end{equation}
to be the noise-weighted sum of the square deviations in power between the signal and the data; $S_\alpha$ are the one-sided power spectral densities of the noise for the frequency bins used
in each corresponding SFT.

A natural way to estimate $\Ap^2$, analogous to $\chi^2$ minimization, is to find the value of $\Ap^2$ that minimizes $g$.  We therefore solve
\begin{equation}
\label{dgdAplus2}
{\partial g \over \partial \Ap^2}
= - \sum_{\alpha} { ( P_\alpha - 0.5 \Ap^2 F_{+ \alpha}^2 \TSFT ) F_{+ \alpha}^2 \TSFT \over S_\alpha^2 }
= 0
\end{equation}
for $\Ap^2$, and obtain
\begin{equation}
\label{APlus}
\Ap^2 =   4 \sum_{\alpha} { F_{+ \alpha}^2 \over S_\alpha^2 }
{| \tilde{x}_\alpha |^2 \over \TSFT^2 }
/
\sum_{\alpha} { F_{+ \alpha}^4 \over S_\alpha^2 } .
\end{equation}

Equation~\eref{APlus} is the detection statistic for the PowerFlux method given in \cite{pf,s4semicoherentpaper}, although the derivation given here is different. We refer to this as the linear PowerFlux method. To instead apply this approach to circularly polarized signals, we replace $F_{+ \alpha}^2$ with $F_{+ \alpha}^2 + F_{\times \alpha}^2$ in equation~\eref{NWSSDP}; this gives what we
refer to as the circular PowerFlux method. The implementation of PowerFlux described in \cite{pf,s4semicoherentpaper} uses the linear PowerFlux method with a search over discrete values of the polarization $\psi$, together with the circular PowerFlux method, to search for elliptically polarized signals.


\section{Generalization to estimate $\Ap^2$ and $\Ax^2$}\label{derivedg}

We now investigate generalizing the above derivation to estimate $\Ap^2$ and $\Ax^2$ simultaneously. The natural generalization of equation~\eref{NWSSDP} is to redefine $g$ as
\begin{equation}
\label{NWSSDP2}
g = \sum_{\alpha} { [ P_\alpha - 0.5 (\Ap^2 F_{+ \alpha}^2 + \Ax^2 F_{\times \alpha}^2) \TSFT ]^2
\over S_\alpha^2 }.
\end{equation}
Following the same minimization procedure, we obtain the equations
\begin{eqnarray}
\label{dg2dAplus2}
{\partial g \over \partial \Ap^2} &=
- \sum_{\alpha} { [ P_\alpha - 0.5 (\Ap^2 F_{+ \alpha}^2 + \Ax^2 F_{\times \alpha}^2) \TSFT ]
F_{+ \alpha}^2 \TSFT \over S_\alpha^2 }
= 0, \\
\label{dg2dAcross2}
{\partial g \over \partial \Ax^2} &=
- \sum_{\alpha} { [ P_\alpha - 0.5 (\Ap^2 F_{+ \alpha}^2 + \Ax^2 F_{\times \alpha}^2) \TSFT ]
F_{\times \alpha}^2 \TSFT \over S_\alpha^2 }
= 0,
\end{eqnarray}
which we solve for $\Ap^2$ and $\Ax^2$ to give
\begin{eqnarray}
\label{Aplus2}
\Ap^2 &= {4 \over {\cal D}} \biggl [
\sum_{\alpha} { F_{\times \alpha}^4 \over S_\alpha^2 }
\sum_{\alpha} { F_{+ \alpha}^2 \over S_\alpha^2 } {| \tilde{x}_\alpha |^2 \over \TSFT^2 }
-
\sum_{\alpha} { F_{+ \alpha}^2 F_{\times \alpha}^2 \over S_\alpha^2 }
\sum_{\alpha} { F_{\times \alpha}^2 \over S_\alpha^2 } {| \tilde{x}_\alpha |^2 \over \TSFT^2 } \biggr ]
, \\
\label{Across2}
\Ax^2 &= {4 \over {\cal D}} \biggl [
\sum_{\alpha} { F_{+ \alpha}^4 \over S_\alpha^2 }
\sum_{\alpha} { F_{\times \alpha}^2 \over S_\alpha^2 } {| \tilde{x}_\alpha |^2 \over \TSFT^2 }
-
\sum_{\alpha} { F_{+ \alpha}^2 F_{\times \alpha}^2 \over S_\alpha^2 }
\sum_{\alpha} { F_{+ \alpha}^2 \over S_\alpha^2 } {| \tilde{x}_\alpha |^2 \over \TSFT^2 } \biggr ]
,
\end{eqnarray}
where
\begin{equation}
\label{calD}
{\cal D} =
\sum_{\alpha} { F_{+ \alpha}^4 \over S_\alpha^2 } \sum_{\alpha} { F_{\times \alpha}^4 \over S_\alpha^2 }
-
\left ( \sum_{\alpha} { F_{+ \alpha}^2 F_{\times \alpha}^2 \over S_\alpha^2 } \right )^2 .
\end{equation}
We refer to this method as generalized PowerFlux I. A natural detection statistic would be $\Ap^2 + \Ax^2$; to evaluate it, we must compute $5/2$ as many summations as for the linear PowerFlux method. We must also, as for linear PowerFlux, still include a search over discrete values of $\psi$.

\section{Generalization to estimate $\Ap^2$, $\Ax^2$, and $\psi$}

To further generalize the PowerFlux method to directly estimate $\psi$, we note that the antenna patterns $F_+$ and $F_\times$ can be written in terms of two functions $a$ and $b$, given
in \cite{jks}, which are independent of $\psi$:
\begin{eqnarray}
\label{Fplus}
F_+ (\psi,t) &= \sin\zeta [ a(t) \cos 2\psi  + b(t) \sin 2\psi ] , \\
\label{Fcross}
F_\times (\psi,t) &= \sin\zeta [ b(t) \cos 2\psi  - a(t) \sin 2\psi ] .
\end{eqnarray}
where $\zeta$ is the angle between the arms of the interferometer. The normalized signal power can therefore be written as
\begin{equation}
\label{pOneSFT3}
{2 | \tilde{h}_\alpha |^2 \over \TSFT } =
0.5 ({\cal A} a_\alpha^2 + {\cal B} b_\alpha^2 + {\cal C} a_\alpha b_\alpha)\TSFT,
\end{equation}
where the amplitudes ${\cal A}$, ${\cal B}$, and ${\cal C}$ are defined to be
\begin{eqnarray}
\label{calA}
{\cal A} &= \sin^2\zeta ( \Ap^2 \cos^2 2\psi  + \Ax^2 \sin^2 2\psi ), \\
\label{calB}
{\cal B} &= \sin^2\zeta ( \Ap^2 \sin^2 2\psi  + \Ax^2 \cos^2 2\psi ), \\
\label{calC}
{\cal C} &= \sin^2\zeta ( \Ap^2 - \Ax^2) 2 \cos 2\psi \sin 2\psi.
\end{eqnarray}
Following the minimization procedure again, we redefine $g$ to be
\begin{equation}
\label{NWSSDP3}
g = \sum_{\alpha} { [ P_\alpha
- 0.5 ({\cal A} a_\alpha^2 + {\cal B} b_\alpha^2 + {\cal C} a_\alpha b_\alpha)\TSFT ]^2
\over S_\alpha^2 } ;
\end{equation}
minimizing $g$ with respect to ${\cal A}$, ${\cal B}$, and ${\cal C}$ gives
\begin{eqnarray}
\label{calAeqn}
{ \partial g \over \partial {\cal A} } &=
- \sum_{\alpha} { [ P_\alpha -
0.5 ({\cal A} a_\alpha^2 + {\cal B} b_\alpha^2 + {\cal C} a_\alpha b_\alpha)\TSFT ] a_\alpha^2 \TSFT
\over S_\alpha^2 }
= 0, \\
\label{calBeqn}
{ \partial g \over \partial {\cal B} } &=
- \sum_{\alpha} { [ P_\alpha -
0.5 ({\cal A} a_\alpha^2 + {\cal B} b_\alpha^2 + {\cal C} a_\alpha b_\alpha)\TSFT ] b_\alpha^2 \TSFT
\over S_\alpha^2 }
= 0, \\
\label{calCeqn}
{ \partial g \over \partial {\cal C} } &=
- \sum_{\alpha} { [ P_\alpha -
0.5 ({\cal A} a_\alpha^2
+ {\cal B} b_\alpha^2 + {\cal C} a_\alpha b_\alpha)\TSFT ] a_\alpha b_\alpha \TSFT
\over S_\alpha^2 }
= 0.
\end{eqnarray}
Thus, the amplitudes ${\cal A}$, ${\cal B}$, and ${\cal C}$ can be found by inverting equations~\eref{calAeqn}-\eref{calCeqn}; the amplitudes $\Ap^2$ and $\Ax^2$ and polarization angle $\psi$ are then found by inverting equations~\eref{calA}-\eref{calC}. This method is referred to as generalized PowerFlux II. It involves computing $8/2=4$ as many sums as the linear PowerFlux method; however, because we no longer need to search over discrete values of $\psi$, the total computational cost of this method may overall be lower than the implementation of PowerFlux described in \cite{s4semicoherentpaper,pf}. An alternate method of estimating $\Ap^2$, $\Ax^2$, and $\psi$, which uses the output of the linear PowerFlux method evaluated at several fixed values of $\psi$, is given in Appendix A of \cite{s4semicoherentpaper}.

\section{Comparison of detection efficiencies}

We have determined the relative detection efficiencies (defined below) of the standard PowerFlux methods, and the generalizations presented above. The detection statistic used for each method is $\Ap^2 + \Ax^2$, with $\Ax^2 = 0$ in the case of the linear PowerFlux method. For simplicity, we did not restrict the estimations of $\Ap^2$ and $\Ax^2$ to the physical region $\Ap^2 \ge 0$ and $\Ax^2 \ge 0$, or check if the actual minimum value of $g$ was on the boundary of this region. An implementation of the StackSlide method described in \cite{s4semicoherentpaper}, which uses a sum of the power as the detection statistic, was also included.

We first performed searches on 10000 sets of SFTs containing only randomly generated Gaussian noise and obtained, for each method, a distribution of their detection statistic in the absence of a signal. From this distribution we determined the threshold for a 1\%~false alarm rate. We then performed searches on 3000 sets of SFTs, each containing noise plus a simulated signal from an isolated spinning source.  In this case the strength of the signal can be given by the normalized injected amplitude $h_0(\TSFT/S)^{1/2}$, where $h_0$ is defined by \cite{jks}
\begin{eqnarray}
\label{Aplusdef}
&\Ap = 0.5 h_0 (1 + \cos^2 \iota) , \\
\label{Acrossdef}
&\Ax = h_0 \cos \iota ,
\end{eqnarray}
and where $\iota$ is inclination angle between the spin axis of the source and the direction from the source to the detector, and $S$ is the one-sided power spectral density of the Gaussian noise. For each of the 3000 simulated signals, $\cos \iota$ was chosen at random from the range $[-1, 1]$, resulting in elliptically polarized signals; the polarization angle $\psi$ was also chosen at random from the range $[-\pi/4, \pi/4]$. 

Other parameters of the signal are its sky position (right ascension and declination) and its frequency at the Solar System Barycenter (SSB), taken as constant here. Since right ascension has minimal effect on the results this was set to zero and results were found for several values of the declination. For the detector configuration we used that of LIGO Hanford, as given in \cite{jks}. We also varied the number of SFTs. 

A real search requires a template bank of sky positions, frequencies (and frequency time derivatives, which we set to zero) at the SSB, and in some cases polarization angles.
Here, for each simulated signal a single template was created with no mismatch in sky position. However, we included a random mismatch between the frequencies of the template and the signal at the SSB of up to half an SFT bin width and, for searches requiring a value for $\psi$, a random mismatch in $\psi$ between the template and the signal of up to $\pi/16$. For each simulated source, the detection statistic was computed using the template sky position, SSB frequency and $\psi$ (if needed); thus included in the results are losses due to mismatch between the template and the signal.

\begin{figure}
\includegraphics[width=2.6in]{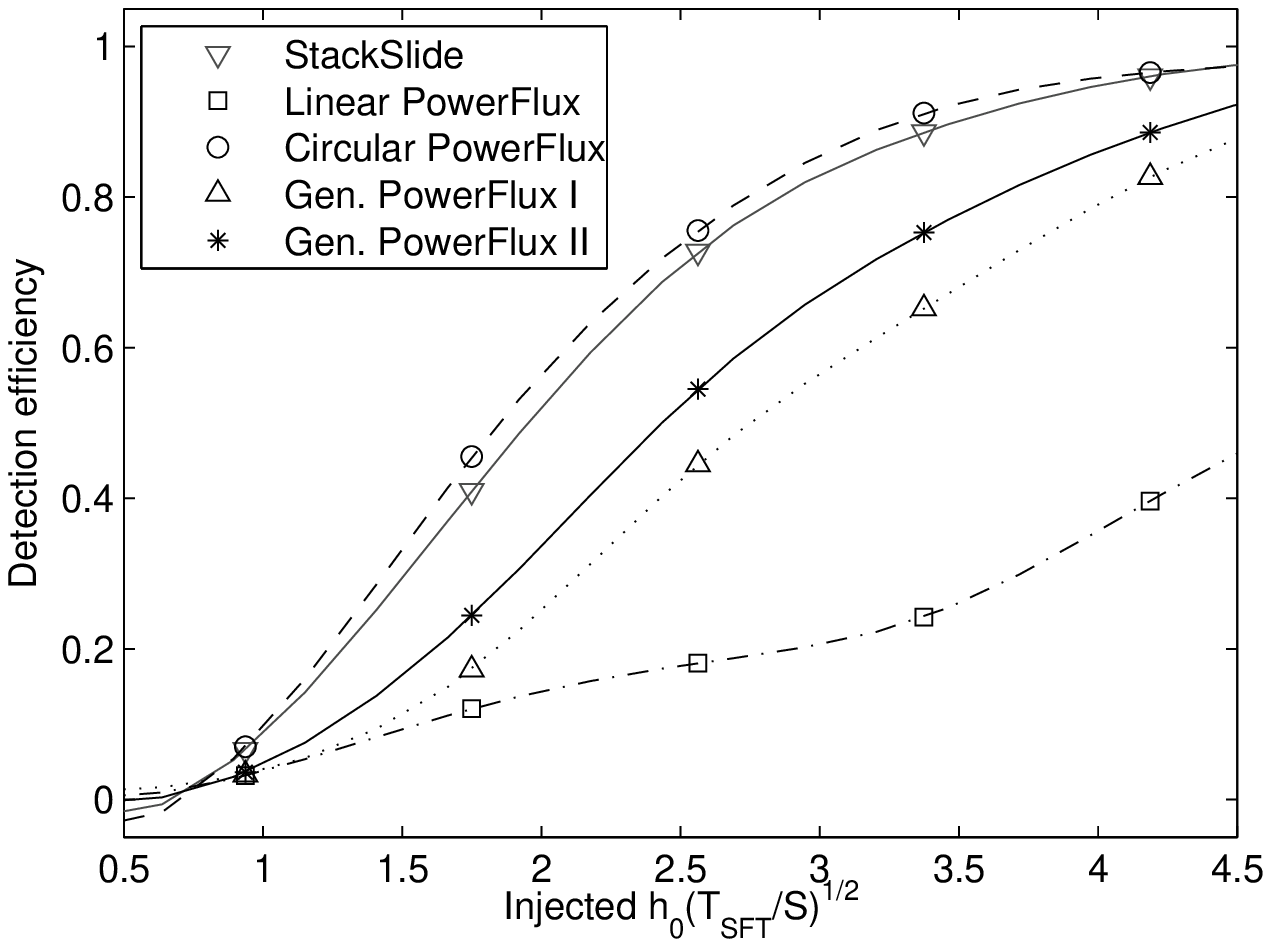}
\includegraphics[width=2.6in]{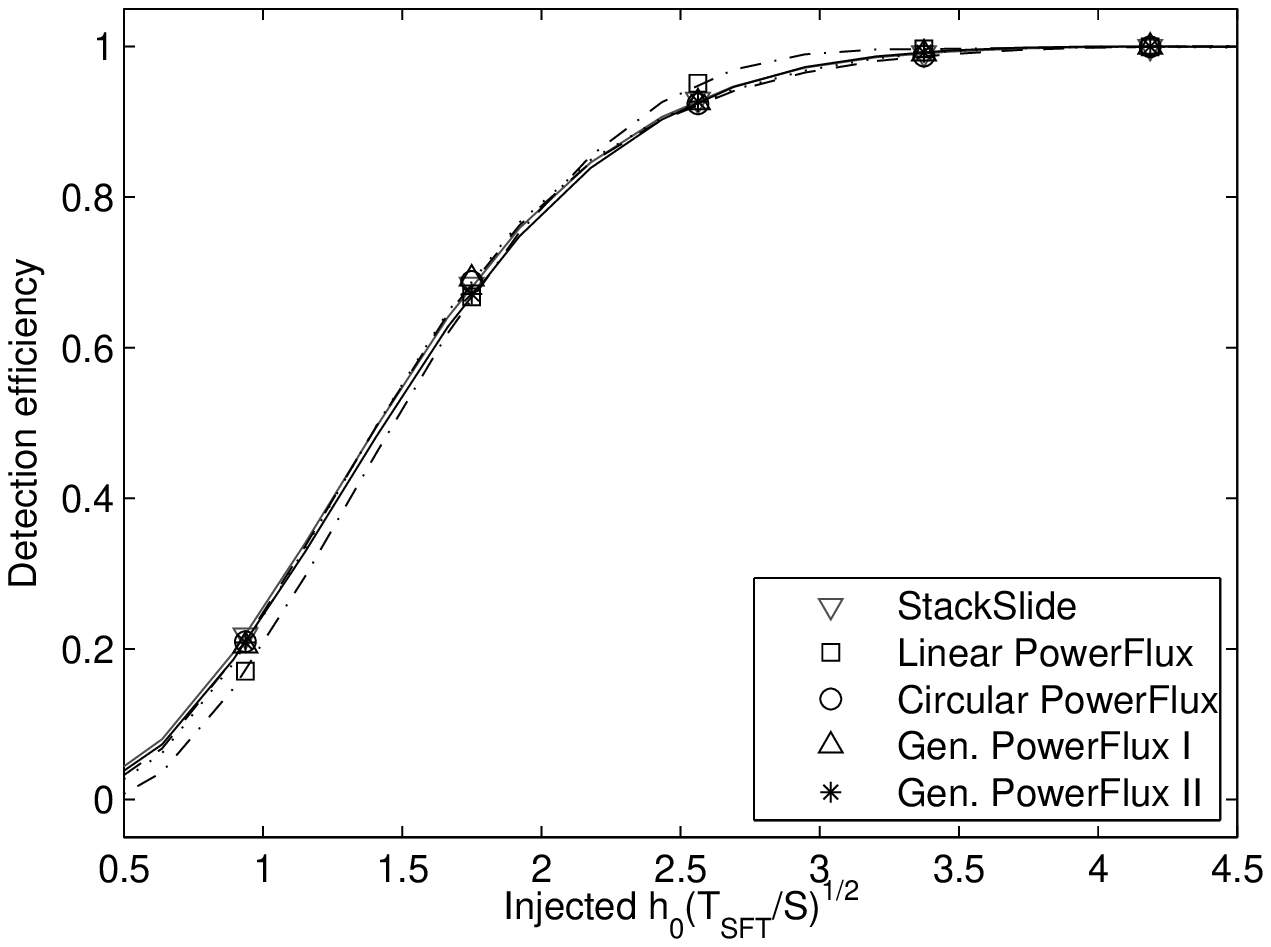}
\caption{\label{Eff1}
Detection efficiency, determined to within 3\% by simulated searches, versus the normalized injected amplitude, for 336 SFTs and signal declinations of zero (left) and 90${}^o$ (right). The marked points are the results of the simulated searches and the solid lines are spline fits to these.}
\end{figure}
\begin{figure}
\includegraphics[width=2.6in]{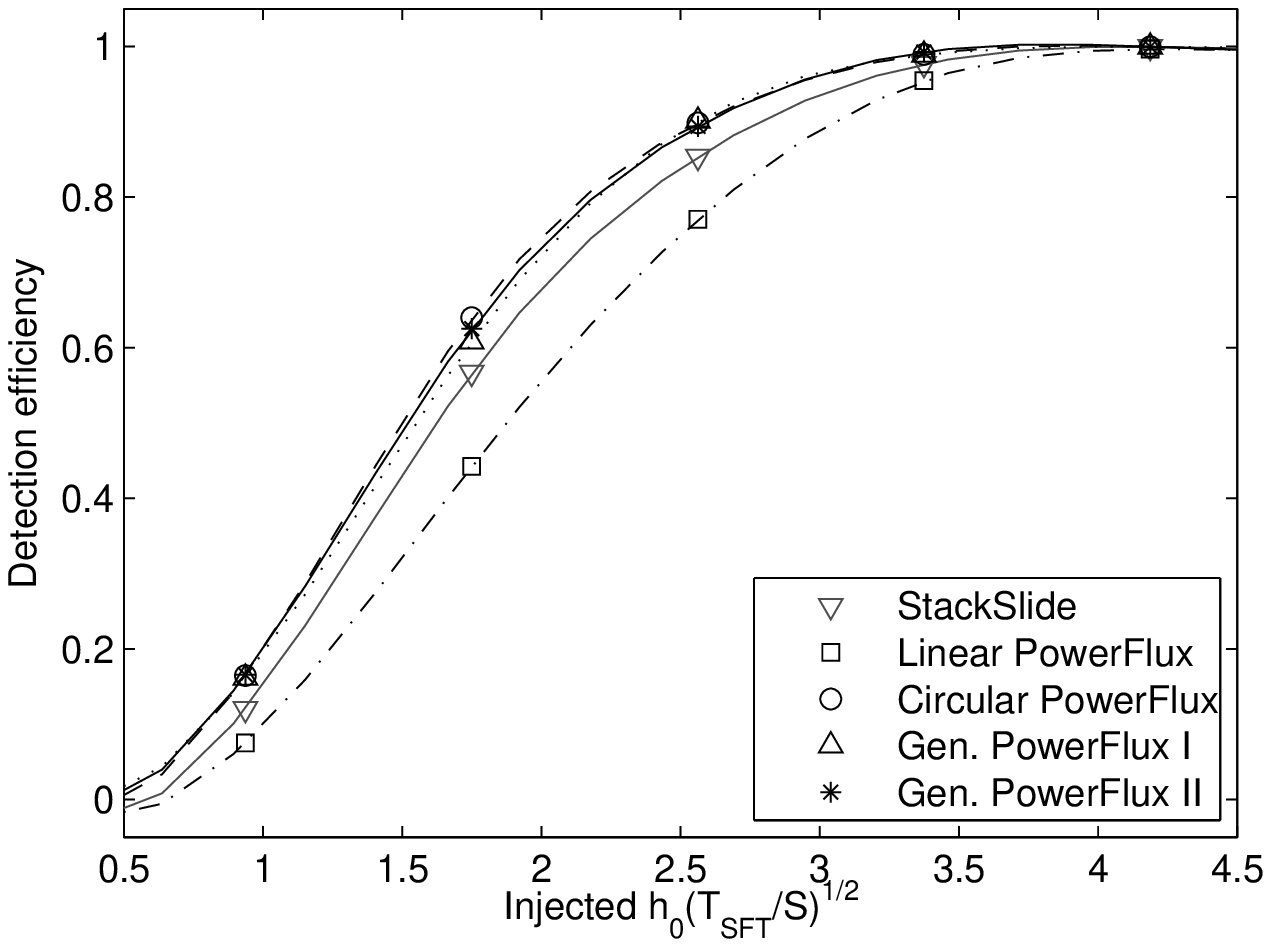}
\includegraphics[width=2.6in]{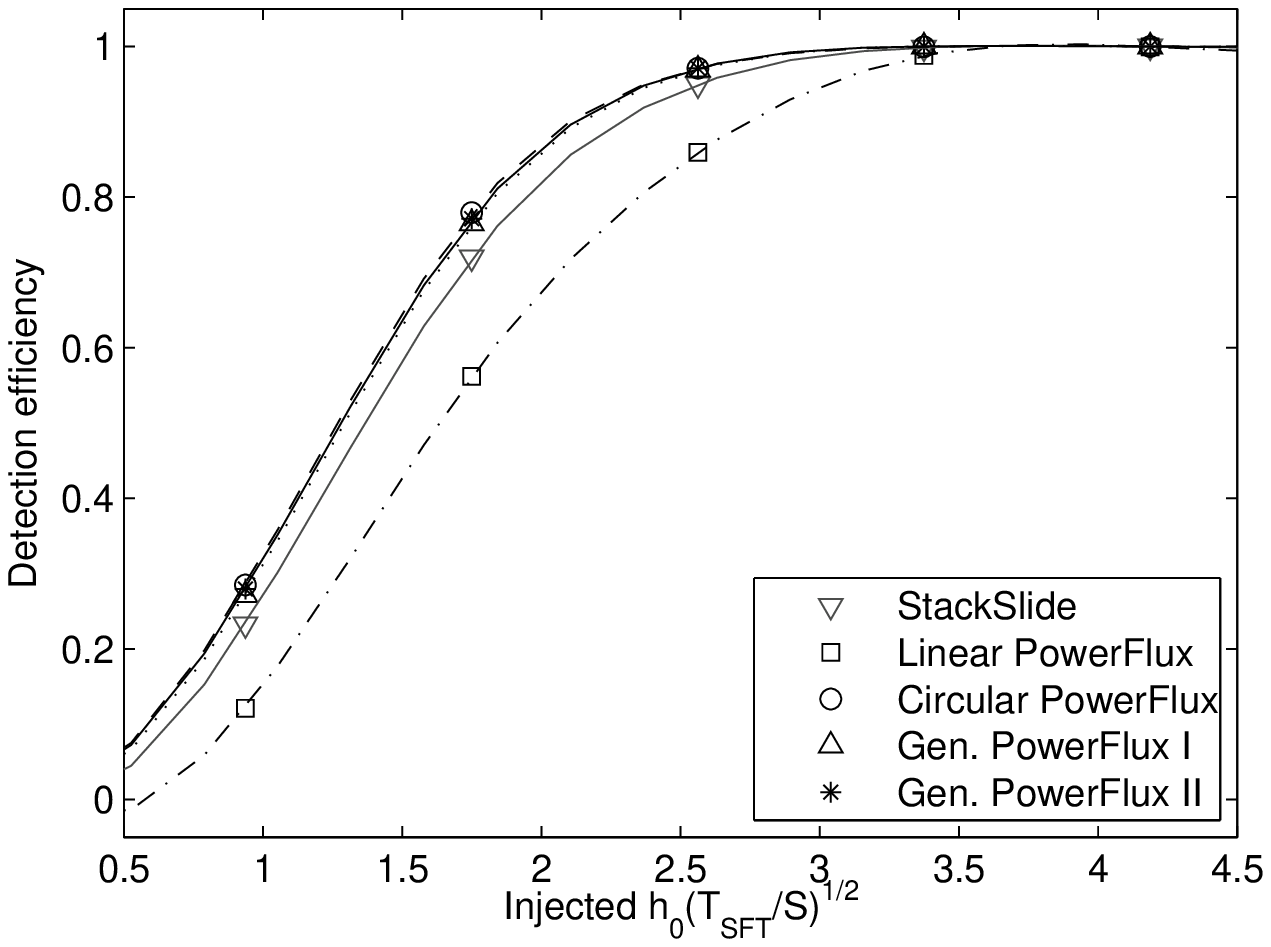}
\caption{\label{Eff2}
Same as figure \ref{Eff1}, for 336 (left) and 672 SFTs (right), and a signal declination of 45${}^o$.}
\end{figure}


After searching the 3000 sets of SFTs we obtained, for each method, a distribution of their detection statistic in the presence of a signal of normalized amplitude $h_0(\TSFT/S)^{1/2}$. The detection efficiency is then calculated as the fraction of the distribution of the detection statistic which falls above the 1\% false alarm rate threshold. We then repeated the process of generating and searching 3000 sets of SFTs for increasing values of the normalized amplitude; this gives the detection efficiency as a function of $h_0(\TSFT/S)^{1/2}$.


To verify our results, the authors independently wrote \textsc{Matlab} scripts to perform the above procedure. The scripts produced identical detection efficiency curves for the same input parameters,
within the expected uncertainties based on the number of searches performed; this gives us confidence that our implementations are correct.

Figures \ref{Eff1} and \ref{Eff2} show the detection efficiencies of the methods versus $h_0(\TSFT/S)^{1/2}$, for a selection of values for the number of SFTs and declinations.
As the source moves away from zero declination, the detection efficiency increases, and the differences
between the methods becomes smaller. Also, the relative efficiencies of StackSlide, the generalized PowerFlux, and the linear PowerFlux methods can change with respect to each other, but note that circular PowerFlux remains typically the most efficient method. This is good news, since this method is already being used by the LSC.
It also may not be surprising, since we note that the weights used in the Hough transform
search in \cite{s4semicoherentpaper} were proportional to $F_+^2 + F_\times^2$.
These weights were shown to be optimal in an average sense in \cite{impHoughJPCS,impHoughTech}, 
from which we see that $\Ap^2 F_+^2 + \Ax^2 F_\times^2$
averaged over $\psi$ can be factored as $0.5(F_+^2 + F_\times^2)(\Ap^2 + \Ax^2)$,
i.e., as the antenna pattern for circular polarization times the sum of the squared amplitudes,
even for an elliptically polarized signal. 
Since our simulations perform a Monte Carlo average over $\psi$, this may explain why circular PowerFlux
does so well.
Finally, note that the detection efficiencies
increase with the number of SFTs, as expected, but the relative efficiencies of the methods
do not change significantly.

\begin{figure}
\includegraphics[width=1.65in]{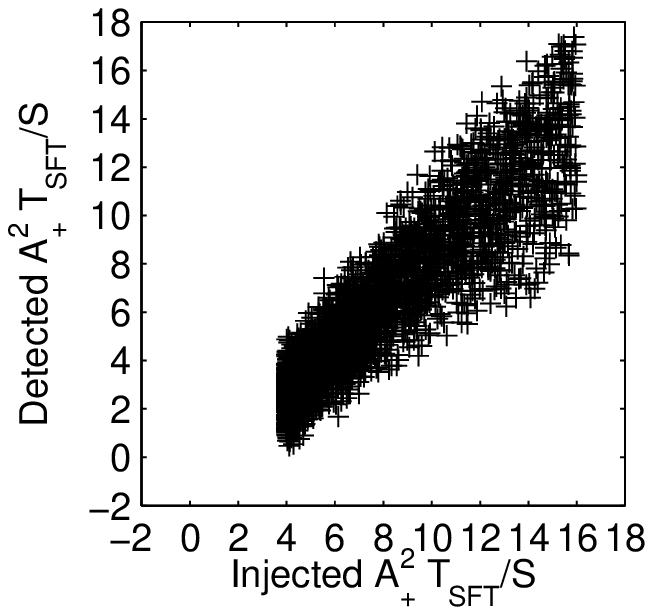}
\includegraphics[width=1.65in]{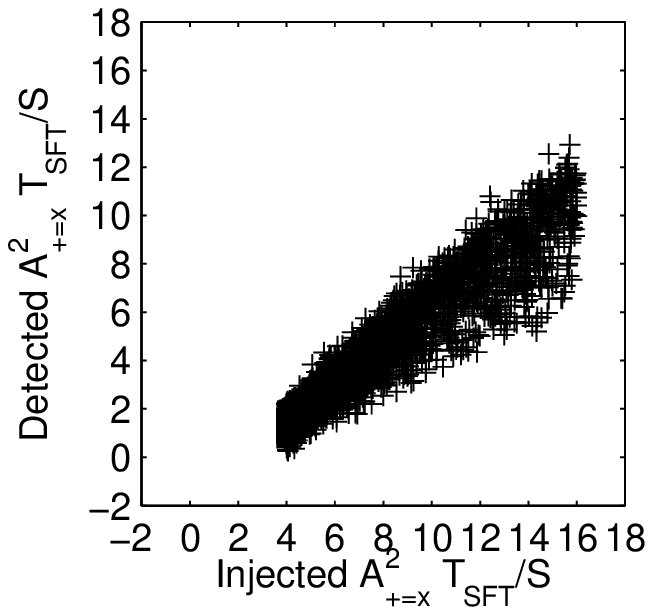}
\includegraphics[width=1.65in]{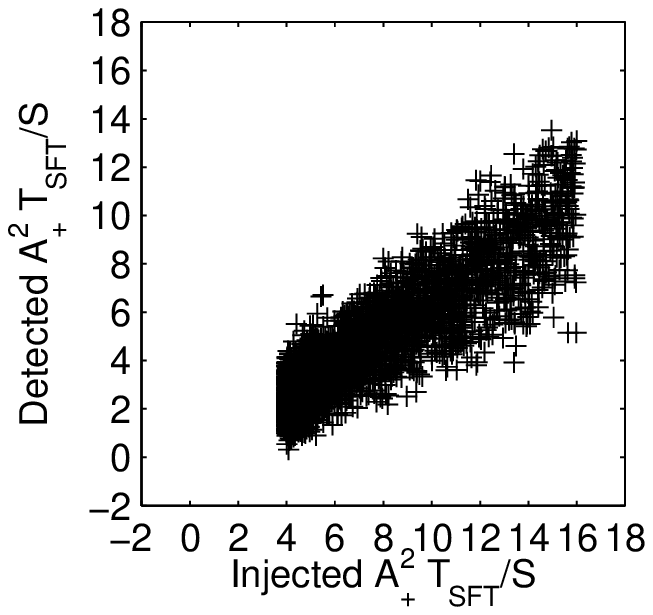}
\includegraphics[width=1.65in]{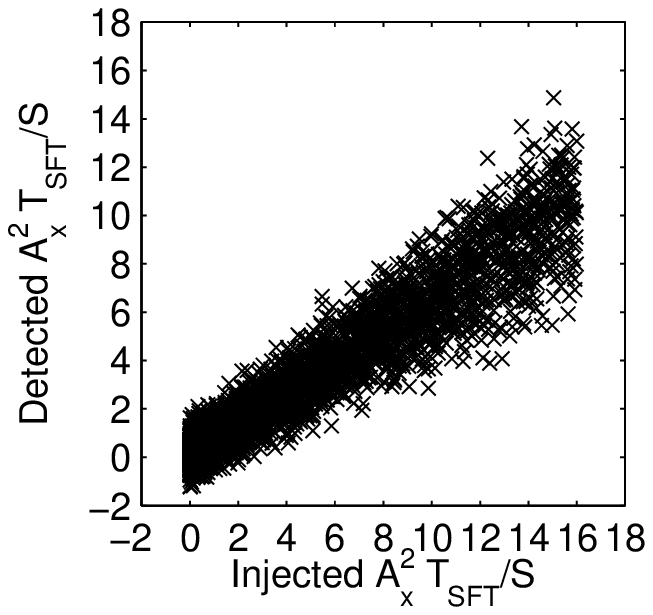}
\includegraphics[width=1.65in]{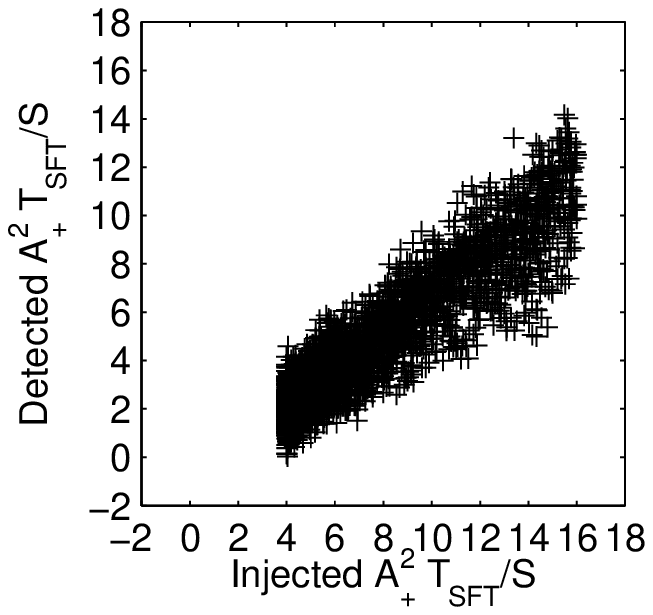}
\includegraphics[width=1.65in]{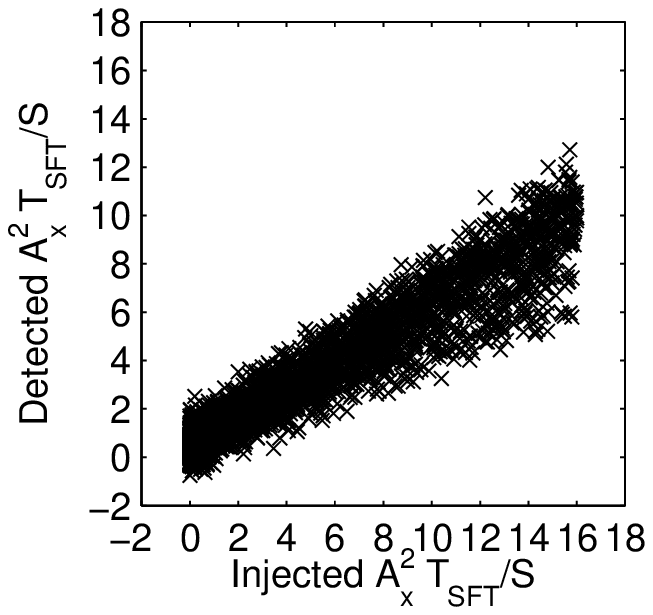}
\caption{\label{Inj1}
Injected vs. detected (after removing the noise only mean) normalized squared amplitudes estimated by (from left to right, top to bottom) linear PowerFlux ($\Ap$), circular PowerFlux ($\Ap = \Ax$), generalized PowerFlux I ($\Ap$ and $\Ax$), and generalized PowerFlux II ($\Ap$ and $\Ax$), for $h_0(\TSFT/S)^{1/2} = 4$.}
\end{figure}
\begin{figure}
\begin{minipage}[b]{3.0in}
\caption{\label{Inj2}
Injected vs. detected polarization angle (in units of $\pi$) estimated by generalized PowerFlux II, for $h_0(\TSFT/S)^{1/2} = 4$.}
\end{minipage}
\includegraphics[width=1.65in]{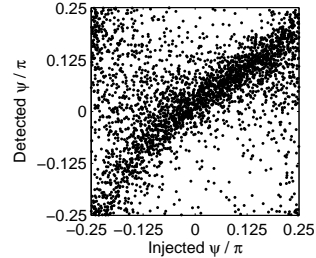}
\end{figure}


Figure \ref{Inj1} shows, for a search of 3000 sets of SFTs injected with $h_0(\TSFT/S)^{1/2} = 4$ signals, the distribution of the injected versus the detected amplitudes estimated by the various methods. The mean value of the squared amplitudes estimated for noise alone was subtracted from the detected 
squared amplitudes; the injected and detected squared amplitudes are then normalized by $\TSFT/S$. The detected squared amplitudes are typically smaller than that injected due to loss of power due to mismatch between the signal and
the template used to find it, while the width of the distributions are due to noise. The generalized PowerFlux II method has marginally the narrowest such distribution. Figure~\ref{Inj2} shows the injected versus the detected polarization angle estimated by generalized PowerFlux II; the estimation can be quite poor, for example when noise causes the detected $\psi$ to show up in the wrong
quadrant.

\section{A PowerFlux maximum likelihood statistic, and future work}

In section \ref{derivedg}, we gave equation~\eref{NWSSDP2} for $g$, in analogy to the $\chi^2$ statistic. A direct comparison fails because $P_\alpha$ is not a Gaussian distributed variable. However, if we expand the square in the numerator, and retain only the terms that depend on $\Ap$ and $\Ax$,
we can define a ``PowerFlux maximum likelihood statistic'', $\mathcal{G}$, by
\begin{equation}
\label{MaxLikelihood}
\fl \qquad \mathcal{G} = \sum_{\alpha} { [ (\Ap^2 F_{+ \alpha}^2 +
\Ax^2 F_{\times \alpha}^2) \TSFT P_\alpha - 0.25
(\Ap^2 F_{+ \alpha}^2 + \Ax^2 F_{\times \alpha}^2)^2 \TSFT^2]
\over S_\alpha^2 },
\end{equation}
where
it is understood
that $\Ap^2$ and $\Ax^2$ are chosen to minimize $g$. For example, we can substitute equation~\eref{APlus} for $\Ap^2$ and set $\Ax^2 = 0$, which gives
\begin{equation}
\label{PFLinMaxLike}
\mathcal{G} =   4 \TSFT^2 \left( \sum_{\alpha} { F_{+ \alpha}^2 \over S_\alpha^2 }
{| \tilde{x}_\alpha |^2 \over \TSFT^2 } \right )^2
/
\sum_{\alpha} { F_{+ \alpha}^4 \over S_\alpha^2 } .
\end{equation}
Here, $\mathcal{G}$ represents our definition of the ``maximum likelihood statistic'' for linear PowerFlux. It is similar to the standard linear PowerFlux statistic given in equation~\eref{APlus}; note, however, that the sum in the numerator is squared. Similar expressions for the circular and generalized PowerFlux methods, and indeed any method that computes either $\Ap^2$ or $\Ax^2$, can be found using equation~\eref{MaxLikelihood}.

For future work we plan to investigate whether using the PowerFlux maximum likelihood
statistic, $\mathcal{G}$, gives a better detection efficiency than the sum of the
squared amplitudes, $\Ap^2 + \Ax^2$, used in this paper. It would also be interesting to further understand
why circular PowerFlux is so efficient (and under precisely which conditions this is so), and study whether
this method or another one is mathematically the optimal filter of SFT power in the Neyman-Pearson sense,
i.e. that maximizes the detection efficiency for a fixed false alarm rate.

\ack
We thank V. Dergachev, B. Krishnan, K. Riles, and A. Sintes for providing useful comments on
a draft of this paper.
G. Mendell gratefully acknowledges the support of Caltech and LIGO Hanford Observatory.
K. Wette is grateful for the support of The ANU, and of Caltech and LIGO Hanford when visiting the observatory. LIGO was constructed by the California Institute of Technology
and Massachusetts Institute of Technology with funding from the National
Science Foundation and operates under cooperative agreement PHY-0107417.
This paper has been assigned LIGO document number LIGO-P070110-00-Z.

\section*{References}
\bibliographystyle{unsrt}
\bibliography{GeneralizedPowerFluxMethods}

\end{document}